\documentstyle[11pt,newpasp,twoside,epsf]{article}
\begin{document}
\title{The Properties of Brightest Cluster Galaxies in X-Ray Selected Clusters}
 \author{Sarah Brough}
\affil{Astrophysics Research Institute, Liverpool John Moores University, England.}
\author{Chris A. Collins}
\affil {Astrophysics Research Institute, Liverpool John Moores University, England.}
\author{Doug J. Burke}
\affil {Harvard-Smithsonian Center for Astrophysics, Cambridge, USA.}
\author{Paul D. Lynam}
\affil {Max-Planck-Institut fur Extraterrestriche Physik, Garching, Germany.}
\author{Robert G. Mann}
\affil {Institute for Astronomy, University of Edinburgh, Edinburgh, England.}

\begin{abstract}
We present the K-band Hubble diagram for 162 brightest cluster galaxies (BCGs) in X-ray selected clusters, $0.01<z<0.83$.  The sample incorporates that of Burke, Collins, \& Mann (2000) and includes additional infrared data from the 2MASS extended source catalogue.  We show that below $z\sim0.1$ the BCGs show no correlation with their environment, however, above $z\sim0.1$ BCGs in more X-ray luminous clusters are more uniform in their photometric properties.  This suggests that there may be two populations of BCGs which have different evolutionary histories.
\end{abstract}

\section{Introduction}
Brightest cluster galaxies (BCGs) provide a unique sample with which to study galaxy evolution in a cluster environment.  In a sample of 78 BCGs, $0.05<z<0.8$, in X-ray selected clusters, Collins \& Mann (1998) and Burke, Collins, \& Mann (2000) observed a split with cluster properties in their K-band Hubble diagram.  This work substantially increases their sample with 84 BCGs below $z\sim0.1$ from the {\it ROSAT\/} All Sky Survey, with $K_{s}$-band data from the 2MASS catalogue.  The colour correction between the 2MASS $K_{s}$-band and the K-band is negligible (Carpenter 2001) and is henceforth neglected.   The 2MASS magnitudes have been checked for consistency against data from the 2m UH telescope.  A full discussion of this work will be presented in Brough et al., in preparation. 

\section{Results and Conclusions}

\begin{figure}
\plotfiddle{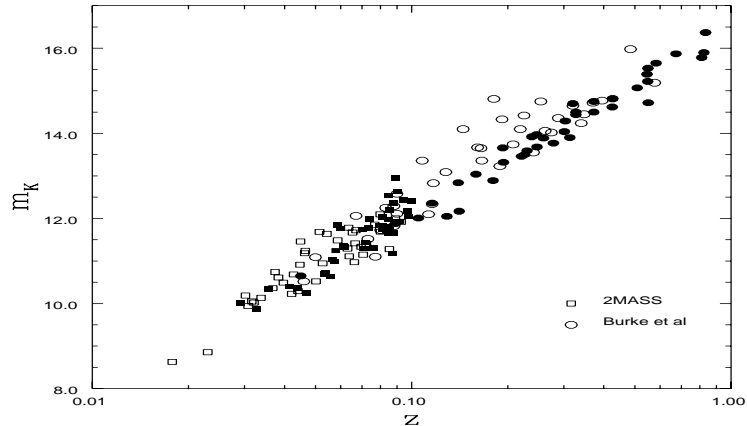}{2in}{0}{50}{30}{-160}{-55}
\caption{The Hubble diagram. The filled points denote BCGs in clusters with $L_{X}(0.3-3.5keV)>2.3\times10^{44}ergs^{-1}$}
\end{figure}

Fig. 1 shows the K-band Hubble diagram for 162 BCGs.  At redshifts beyond $z>0.1$ BCG magnitudes depend on the X-ray luminosity of the host environment, as suggested by Collins \& Mann (1998).  In particular, clusters with $L_{X}(0.3-3.5keV)>2.3\times10^{44}ergs^{-1}$ have a scatter of 0.24 mag, compared to those in the low luminosity clusters which have an rms dispersion of 0.5 mag and a Kolmogorov-Smirnov test shows that BCGs in the different environments are different at $>99.9$ per cent significance.  However, the 97 clusters below z=0.1 indicate that locally BCG magnitudes have a dispersion of 0.3 mag and are uncorrelated with the X-ray luminosity of their host cluster.  A KS test shows that they are drawn from the same population at 90 per cent level.  These differences suggest that there may be two populations of BCGs with different evolutionary histories: BCGs in less X-ray luminous clusters evolving through a process of mergers and those in more luminous clusters evolving passively.  This result has also been seen by Zaritsky et al. in this meeting (astro-ph/0108152).

\section*{Acknowledgments}
This publication makes use of data products from 2MASS which is a joint project of the University of Massachusetts and the Infrared Processing and Analysis Center/California Institute of Technology, funded by the National Aeronautics and Space Administration and the National Science Foundation.

\end{document}